\documentclass[acmsmall]{acmart}
\usepackage{packages}

\setcopyright{cc}
\setcctype{by}

\acmDOI{10.1145/3730985}
\acmYear{2025}
\acmVolume{3}
\acmNumber{CoNEXT2}
\acmArticle{13}
\acmMonth{6}
\acmJournal{PACMNET}

\received{December 2024}
\received[revised]{April 2025}
\received[accepted]{April 2025}

\begin{document}

\setlength{\TPHorizModule}{\paperwidth}
\setlength{\TPVertModule}{\paperheight}
\begin{textblock}{0.802}(-0.025,-0.09)
  \fbox{\begin{minipage}{\linewidth}
    \noindent
    \footnotesize
    If you cite this paper, please use the ACM CoNEXT reference: Marcel Kempf, Simon Tietz, Benedikt Jaeger, Johannes Späth, Georg Carle, and Johannes Zirngibl. 2025. QUIC Steps: Evaluating Pacing Strategies in QUIC Implementations. Proc. ACM Netw. 3, CoNEXT2, Article 13 (June 2025), 14 pages. https://doi.org/10.1145/3730985
  \end{minipage}}
\end{textblock}

\title{QUIC Steps: Evaluating Pacing Strategies in QUIC Implementations}

\author{Marcel Kempf}
\email{kempfm@net.in.tum.de}
\orcid{0000-0001-7540-776X}
\affiliation{%
  \institution{Technical University of Munich}
  \city{Munich}
  \country{Germany}
}

\author{Simon Tietz}
\email{tietz@net.in.tum.de}
\orcid{0009-0007-9470-463X}
\affiliation{%
  \institution{Technical University of Munich}
  \city{Munich}
  \country{Germany}
}

\author{Benedikt Jaeger}
\email{jaeger@net.in.tum.de}
\orcid{0000-0002-8541-5496}
\affiliation{%
  \institution{Technical University of Munich}
  \city{Munich}
  \country{Germany}
}

\author{Johannes Späth}
\email{spaethj@net.in.tum.de}
\orcid{0009-0008-5564-2090}
\affiliation{%
  \institution{Technical University of Munich}
  \city{Munich}
  \country{Germany}
}

\author{Georg Carle}
\email{carle@net.in.tum.de}
\orcid{0000-0002-2347-1839}
\affiliation{%
  \institution{Technical University of Munich}
  \city{Munich}
  \country{Germany}
}

\author{Johannes Zirngibl}
\email{jzirngib@mpi-inf.mpg.de}
\orcid{0000-0002-2918-016X}
\affiliation{%
  \institution{Max Planck Institute for Informatics}
  \city{Saarbrücken}
  \country{Germany}
}

\renewcommand{\shortauthors}{Marcel Kempf et al.}

\begin{abstract}

Pacing is a key mechanism in modern transport protocols, used to regulate packet transmission timing to minimize traffic burstiness, lower latency, and reduce packet loss.
Standardized in 2021, \quic is a \acs{udp}-based protocol designed to improve upon the \acs{tcp}\,/\,\acs{tls} stack.
While the \quic protocol recommends pacing, and congestion control algorithms like BBR rely on it, the user-space nature of \quic introduces unique challenges.
These challenges include coarse-grained timers, system call overhead, and \acs{os} scheduling delays, all of which complicate precise packet pacing.

This paper investigates how pacing is implemented differently across \quic stacks, including \quiche, \picoquic, and \ngtcp, and evaluates the impact of system-level features like \acs{gso} and Linux \acsp{qdisc} on pacing.
Using a custom measurement framework and a passive optical fiber tap, we establish a baseline with default settings and systematically explore the effects of qdiscs, hardware offloading using the ETF qdisc, and \acs{gso} on pacing precision and network performance.
We also extend and evaluate a kernel patch to enable pacing of individual packets within \acs{gso} buffers, combining batching efficiency with precise pacing.
Kernel-assisted and purely user-space pacing approaches are compared.
We show that pacing with only user-space timers can work well, as demonstrated by \picoquic with BBR.

With \quiche, we identify \acs{fq} as a \acs{qdisc} well-suited for pacing \quic traffic, as it is relatively easy to use and offers precise pacing based on packet timestamps.
We uncovered that internal mechanisms, such as a library's spurious loss detection logic or algorithms such as HyStart++, can interfere with pacing and cause issues like unstable congestion windows and increased packet loss.
Our findings provide new insights into the trade-offs involved in implementing pacing in \quic and highlight potential optimizations for real-world applications like video streaming and video calls. %

\end{abstract}



\maketitle

\acresetall

\clearpage

\section{Introduction}
\label{sec:introduction}

\quic has rapidly gained adoption and now handles a significant share of global network traffic~\cite{facebook-quic, cloudflare-adoption, zirngibl2021over9000, zirngibl2023quic}.
It is used for applications like web browsing, video streaming, and video calls.
As \quic continues to grow further in importance, improving its performance is an active area of research.

Pacing, the controlled scheduling of packet transmissions, is a well-known mechanism for improving efficiency by trying to avoid bursty packet transmissions.
It is known to reduce packet loss, lower queuing delays, and improve overall network performance by spacing out packets more evenly over time~\cite{de2018optimizing,zhang2024quickenough}. %
The \quic standard issues the normative statement that a sender should pace outgoing traffic with, for example, a leaky bucket algorithm~\cite{rfc9002}.
As \acp{cca} like \ac{bbr} depend on pacing to work, most \quic implementations contain at least some form of pacing.

However, implementing pacing in \quic presents specific challenges.
Unlike \acs{tcp}, which is paced in the kernel, \quic runs in user-space.
Therefore, \quic stacks must implement pacing themselves, which may not be as effective as kernel-based pacing due to latency and jitter added by system calls and \ac{os} scheduling.
Some implementations rely on kernel features, especially Linux \acp{qdisc}, to help with pacing.
Additionally, techniques like \ac{gso}, which are designed to improve performance, might interfere with pacing by releasing bursts of packets together, counteracting the smoothing effect of pacing.

Since QUIC as a general purpose transport protocol is used for video streaming and video calls, understanding how to implement effective pacing is important for these real-world applications.
This paper explores how pacing is implemented in different \quic stacks, examines the impact of \ac{gso}, and investigates whether qdiscs can improve \quic pacing.

\vspace{0.8em}
\noindent
This work covers the following contributions:

\begin{itemize}[topsep=5pt,leftmargin=24pt]
    \item[\first] We evaluate the pacing behavior of \quic libraries on the wire and find different behavior between libraries and configurations. While \picoquic shows the largest bursts with loss-based \acp{cca}, its pacing behavior using \ac{bbr} outperforms other implementations without using kernel functionalities. 
    \item[\second] We evaluate different mechanisms with and without kernel support, \eg \ac{fq} or \ac{gso} and their impact on the pacing behavior. Existing mechanisms offer \quic flexibility, but an informed optimization of \quic implementations and used mechanisms is important to reach optimal behavior for individual applications and the network in general.
    \item[\third] We extend and evaluate a kernel patch that enables pacing within \ac{gso} buffers. This allows \quic stacks to retain the performance benefits of batching without sacrificing pacing.
    \item[\fourth] We provide all shown patches, configurations, measurement results, evaluation scripts, and our measurement framework to allow others to evaluate \quic pacing in the future and optimize their applications. All data is available in a public GitHub repository~\cite{quic-pacing-paper-repo}.
\end{itemize}
\vspace{0.5em}
We introduce important background in \Cref{sec:background}, followed by a description of our measurement framework, used implementations and features in \Cref{sec:approach}.
\Cref{sec:evaluation} evaluates the behavior of different \quic libraries in combination with system functionalities.
Finally, we cover related work in \Cref{sec:related-work} and conclude our findings in \Cref{sec:conclusion}.

\section{Background}
\label{sec:background}

This section provides relevant background for \quic and pacing, as well as the Linux kernel features that are used in this work.

Pacing denotes the even distribution of packets over a \ac{rtt}~\cite{Aggarwal2000pacing}.
The goal of pacing is to reduce queuing and hereby packet loss by avoiding bursty packet transmissions.

Pacing is integrated in the Linux kernel for more than a decade~\cite{kernel-fq}.
However, \ac{tcp} traffic is only paced when either the selected congestion control algorithm performs pacing or a \ac{qdisc} is used that supports pacing.
For example, \bookworm uses \acs{tcp} \cubic and \ac{fqcodel} as defaults~\cite{bookworm}, thus \ac{tcp} traffic is not paced.
Regarding \quic, RFC 9002~\cite{rfc9002} states that packet bursts have to be prevented and suggests pacing in combination with the used \ac{cca}.
During the current discussions to define a variable \ac{ack} frequency~\cite{draft-ietf-quic-ack-frequency}, the suggestion to pace is further highlighted.
While a smaller \ac{ack} frequency reduces the overhead for data receivers, it reduces the effectiveness of \ac{ack}-clocking and could lead to bursts if pacing is not implemented.

\quic traffic can be paced by a \ac{qdisc}, a mechanism connecting the Linux kernel's networking stack and a \ac{nic}.
It queues and schedules packets for transmission.
\acp{qdisc} can prioritize, delay, or drop packets allowing features like rate limiting, traffic shaping, or pacing.
\quic traffic profits from a \ac{qdisc} that supports the scheduling of packets based on a timestamp, which can be passed to the kernel for every packet. 
Pacing approaches with other \acp{qdisc} involving rate limiting, \eg \ac{tbf}, do not offer an interface for adjusting the pacing rate dynamically from user-space.
They are therefore less interesting for the usage with \quic, as the desired pacing rate can change continuously during a connection.
We look at the \ac{fq} and \ac{etf} \acp{qdisc} in \Cref{sec:evaluation}, as both of them support the previously mentioned timestamp-based scheduling.
By setting the \sotxtime socket option, a timestamp can be passed with each \texttt{sendmsg} call using the \scmtxtime control message header.
The main difference between \ac{fq} and \ac{etf} is that \ac{etf} drops packets if their timestamp is in the past, while \ac{fq} does not.
To ensure that packets are enqueued before their scheduled transmission time and thus not dropped, \ac{etf} has a \textit{delta} parameter that specifies a time offset at which the \ac{qdisc} becomes active in advance~\cite{tc-etf}.
The optimal value for this parameter depends on the processing overhead and varies from system to system.
However, \ac{etf} supports hardware offloading for \acp{nic} that support the so-called \launchtime feature.
In this case, the \ac{nic} holds back outgoing packets until the timestamp is reached, sending them out at the specified timestamp.
This could further increase the precision of pacing (see \Cref{sec:evaluation-etf}).

To reduce the overhead of context switches between \quic in user-space and the kernel, segmentation offloading can be used.
This technique allows passing multiple packets between the kernel network stack and the user space as one data unit~\cite{kerneloffload}.
\ac{gso} and \acl{gro} can be used with \quic and can significantly improve the throughput~\cite{KEMPF202490}.
However, \ac{gso} prevents pacing within each batch of packets.
We focus on the sending behavior and thus \ac{gso} (see \Cref{sec:evaluation-gso}).

\section{Approach}
\label{sec:approach}

This section describes our measurement setup and the developed framework for reproducible measurements.
We also introduce the network emulation techniques to achieve realistic conditions and a controlled bottleneck buffer.
Finally, the tested \quic implementations and their pacing behavior are shortly presented.

\subsection{Measurement Framework}
\label{sec:approach-framework}

To perform reproducible measurements, we use the framework introduced by \citet{jaeger2023quic}.
The framework is designed to set up and configure measurements on bare-metal servers, offering a consistent environment for QUIC experiments with configurable and extensible logging options.
It automates two hosts as client and server to evaluate properties of single QUIC connections in different scenarios.
The server is set up to be a QUIC server hosting a file of flexible size.
The client is instructed to download this file.
No information for session resumption is stored on the client, so that each repetition has the same conditions.
We focus on the server’s pacing behavior during file transmission, as this is the direction of continuous user data flow.

\begin{figure}
	\centering
	\includegraphics[width=0.95\linewidth]{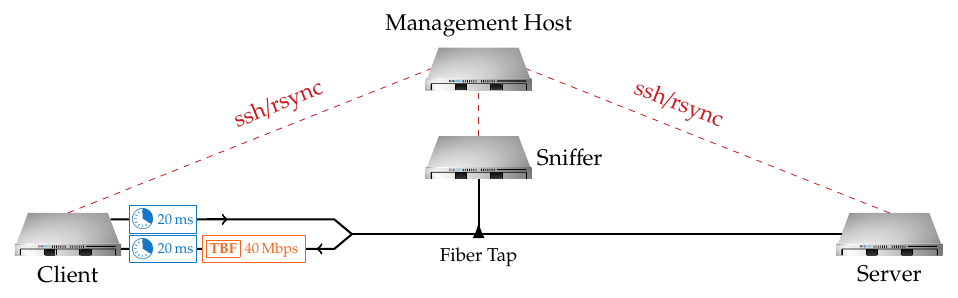}
	\caption{Topology of the measurement setup with an optical fiber tap. Dashed links only carry management traffic, solid links only carry measurement traffic. Delay and rate limiting are denoted in blue and orange.}
	\Description[Topology of the measurement setup.]{Topology of the measurement setup with an optical fiber tap between client and server.} %
    \label{fig:testbed-splitter-setup}
\end{figure}

To evaluate the pacing behavior and inter-packet gap, precise timestamps are required.
When capturing timestamps at the server, the capturing itself might influence the timing of the packets.
On the other side, capturing timestamps on the client only provides accurate values if no network emulation is applied, as network emulation usually re-shapes traffic.
To avoid these issues, we extended the framework to support a third measurement host, the sniffer, which is responsible for capturing packets on the wire between client and server.
We rely on a passive optical fiber tap, which forwards a copy of all packets to the sniffer.
The resulting topology is shown in \Cref{fig:testbed-splitter-setup}.
The optical fiber tap and the sniffer allow us to capture precise timing information without influencing the connection or involved hosts.
\textsl{MoonGen}~\cite{moongen-imc2015} is used to capture packets on the sniffer as it offers high precision with timestamp resolutions below \SI{2}{\nano\second}.

Client and server are both equipped with an \textit{Intel I210 Gigabit Fiber} \ac{nic}.
We chose this card, as it provides support for the \launchtime feature, which is part of our measurements.
The sniffer is equipped with an \textit{Intel E810-XXV} \ac{nic}.
All hosts run \bookworm with client and server running on Linux kernel \textsl{6.1.112-rt30} and the sniffer running on \textsl{6.1.0-17-amd64}.

\subsection{Network Emulation}
\label{sec:approach-netem}

In our setup, client and server are directly connected by a \SI{1}{\Gbps} link with a latency of less than \SI{1}{\milli\second}.
As this ultra low \ac{rtt} might cause unwanted side effects in the congestion control mechanisms, we emulate a bandwidth of \SI{40}{\Mbps} with a minimum \ac{rtt} of \SI{40}{\milli\second}.
While real-world networks may introduce additional dynamics such as cross traffic, queue sharing, competing flows, or load balancing, we intentionally avoid these complexities to ensure reproducibility and focus on stack behavior.
We use \textsl{tc} to set up a \ac{tbf} for bandwidth limitation and \textsl{netem} for the \ac{rtt} emulation~\cite{tbf,netem}.
While \acp{qdisc} are usually used for egress traffic, we need to shape the incoming traffic at the client to not influence the pacing done by the server sending data.
Therefore, we use an intermediate functional block on the client to use a \ac{tbf} on incoming traffic at the client.
The bandwidth from client to server does not need to be limited as mainly \acp{ack} are sent in this direction.
To achieve the desired \ac{rtt}, \SI{20}{\milli\second} are applied directly after the \ac{tbf} as well as for outgoing traffic at the client.
This split is necessary to avoid measuring the uncommon scenario of highly asymmetric latency.
However, to avoid unwanted packet loss due to the usage of \textsl{netem} after \ac{tbf}, we increase the buffer size to fit two \aclp{bdp} of packets.
Due to the independent monitoring with the sniffer, we are able to capture packets sent by the server before any traffic shaping is done.

\subsection{Implementations}
\label{sec:approach-impls}

For our measurements, we use the example client\,/\,server implementations of the \quic libraries \quiche~\cite{quiche}, \picoquic~\cite{picoquic}, and \ngtcp~\cite{ngtcp2}.
We chose these implementations as they follow different approaches for pacing.
While the pacing rate calculation works the same for all three implementations, the actual pacing is done differently.
Cloudflare's \quiche calculates an optimal sending time for each packet and uses the \sotxtime socket option as described in \Cref{sec:background} to pass this timestamp to the kernel.
A \ac{qdisc} like \ac{fq} or \ac{etf} can then schedule the packets based on this timestamp.
In contrast, \ngtcp does not depend on system clocks.
An application is responsible to send packets at the correct time as calculated by the library.
For both implementations, the timestamp of the current packet is based on the previous packet's timestamp and the pacing rate.
\picoquic depends on the application to respect timestamps and wait until it is allowed to send, but it uses a leaky bucket algorithm for pacing as proposed in RFC 9002~\cite{rfc9002}.
This credit-based approach allows small bursts after inactivity in contrast to the interval-based approach of \quiche and \ngtcp.
For the comparison with \tcptls, we use \textsl{nginx} and \textsl{wget} explicitly enabling TLS to ensure a fair comparison with \quic, which includes both transport and encryption~\cite{kempf2024quiccrypto}.

\subsection{Limitations}

The results presented in \Cref{sec:evaluation} are based on a single, specific network configuration (\SI{40}{\Mbps} bandwidth, \SI{40}{\milli\second} minimum \ac{rtt}).
We chose these conditions to study pacing in a controlled environment.
This setup allows for reproducible measurements and focuses on stack behavior.
The exact findings are specific to these fixed parameters and may not be directly generalizable to all network conditions
However, general trends and differences in behavior are visible and explainable with the implementations.
Real networks have different and complex properties, like varying burstiness, cross traffic, shared queues, competing connections, or load balancing.
The impact of pacing and burstiness can be highly context-dependent, and it can be expected that the results vary under different network configurations.
We leave the evaluation of pacing in further network scenarios to future work.

\subsection{Ethics}

This work relies solely on a controlled test environment utilizing synthetic data.
Therefore, it does not involve any real-world user traffic or personal information, raising no privacy concerns or other ethical issues.

\section{Evaluation}
\label{sec:evaluation}

We evaluate the pacing behavior of \quic implementations using the measurement framework described in \Cref{sec:approach}.
Our goal is to analyze how different configurations and system-level features influence pacing.
We establish a baseline using \quic implementations and the \ac{os} in their default configurations.
We compare results to the \tcptls stack.
This provides insights into the general behavior of libraries and allows a comparison with more advanced configurations.
Next, we investigate the impact of different Linux \acp{qdisc} on pacing, focusing on their ability to regulate packet transmission.
Subsequently, we analyze how \ac{gso} affects pacing in \quic, also considering an approach that paces packets from a single \ac{gso} buffer inside the kernel.
Finally, we evaluate hardware offloading capabilities by enabling the \ac{etf} \ac{qdisc} with support for \launchtime.

During each measurement, the server transferred a \SI{100}{\mebi\byte} file to the client over \acs{http} (v3 for \quic and v2 for TCP/TLS).
Each configuration was repeated 20 times and the values of all repetitions are combined for the evaluation.
Before combining measurements, we verified the stability of results and found that the presented inter-packet gap and packet train length metrics showed a small standard deviation.
The UDP receive buffer size was increased to \SI{50}{\mebi\byte} to prevent packet loss at the client~\cite{KEMPF202490}.

\begin{figure}
    \begin{minipage}{0.48\linewidth}%
        \centering
        \includegraphics[width=\linewidth]{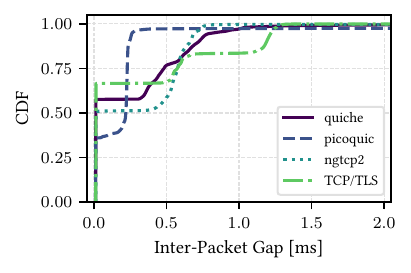}%
        \captionof{figure}{Distribution of gaps between packets sent by the server during the download for the baseline measurements. All implementations use \cubic.\\}%
        \label{fig:baseline-iat-cdf}%
    \end{minipage}%
    \hfill%
    \begin{minipage}{0.48\linewidth}%
        \centering
        \includegraphics[width=\linewidth]{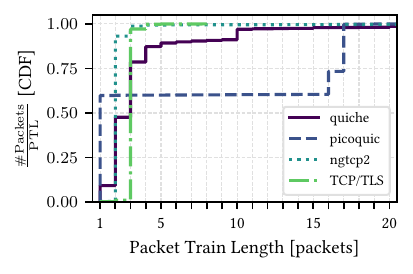}%
        \captionof{figure}{Distribution of packets across packet trains with specific lengths (x-axis) for the baseline. Packet trains are sequences of packets with at most \SI{0.1}{\milli\second} between two packets.}%
        \label{fig:baseline-trainsize-pp-cdf}%
    \end{minipage}%
\end{figure}

\subsection{Baseline Measurement}
\label{sec:baseline}

\newcommand{\resultsBaselineQuicheDrops}{687.15}
\newcommand{\resultsBaselineQuicheDropsStd}{338.12}
\newcommand{\resultsBaselineQuicheGoodput}{34.67}
\newcommand{\resultsBaselineQuicheGoodputStd}{0.64}
\newcommand{\resultsBaselinePicoquicDrops}{861.45}
\newcommand{\resultsBaselinePicoquicDropsStd}{99.53}
\newcommand{\resultsBaselinePicoquicGoodput}{37.09}
\newcommand{\resultsBaselinePicoquicGoodputStd}{0.03}
\newcommand{\resultsBaselineNgtcpDrops}{503.45}
\newcommand{\resultsBaselineNgtcpDropsStd}{7.39}
\newcommand{\resultsBaselineNgtcpGoodput}{15.93}
\newcommand{\resultsBaselineNgtcpGoodputStd}{0.00}
\newcommand{\resultsBaselineTcptlsDrops}{16.50}
\newcommand{\resultsBaselineTcptlsDropsStd}{0.67}
\newcommand{\resultsBaselineTcptlsGoodput}{37.37}
\newcommand{\resultsBaselineTcptlsGoodputStd}{0.02}

For the baseline measurements, we use the default settings for all \quic implementations and set the \ac{cca} to \cubic for comparability.
\Cref{fig:baseline-iat-cdf} shows the \ac{cdf} of the inter-packet gaps.
We observe that even though the distributions differ, approximately \SI{50}{\percent} of the packets are sent back-to-back without being paced.
\picoquic sends slightly fewer packets back-to-back, only reaching \SI{40}{\percent}.
Another commonality is that the majority of packets is sent with inter-packet gaps of less than \SI{1.5}{\milli\second}.

Looking only at these distributions does not provide a clear picture of the pacing behavior.
To gain further insights, we analyze the distribution of the lengths of the packet trains in \Cref{fig:baseline-trainsize-pp-cdf}.
All consecutive packets with an inter-packet gap of \textless\SI{0.1}{\milli\second} each are considered a packet train.
Since the minimum theoretical inter-packet gap in our setup is approximately \SI{0.012}{\milli\second}, a \SI{0.1}{\milli\second} threshold is sufficiently larger to distinguish between fundamental serialization delays and actual bursts or intentionally paced packets.
A packet train of size one is a single packet.
As larger packet trains are bursts, we consider small packet trains and a rather constant inter-packet gap to be necessary for good pacing.

\captionsetup[subfigure]{font=small,skip=0pt}

\begin{figure*}[t]
	\begin{subfigure}{0.33\linewidth}
		\centering
		\includegraphics[width=\linewidth]{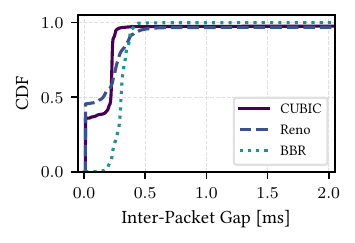}
	\end{subfigure}%
	\begin{subfigure}{0.33\linewidth}
		\centering
		\includegraphics[width=\linewidth]{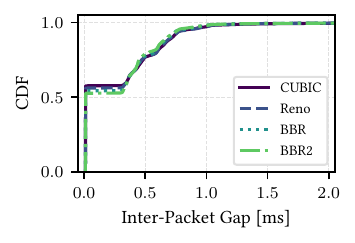}
	\end{subfigure}%
	\begin{subfigure}{0.33\linewidth}
		\centering
		\includegraphics[width=\linewidth]{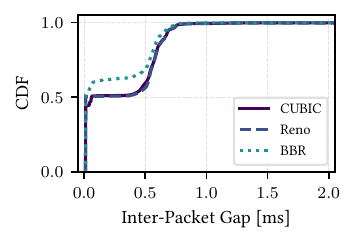}
	\end{subfigure}\\
	\begin{subfigure}{0.33\linewidth}
		\centering
		\includegraphics[width=\linewidth]{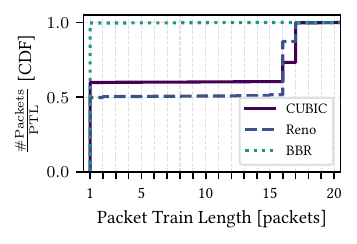}%
		\caption{\picoquic}
	\end{subfigure}%
	\begin{subfigure}{0.33\linewidth}
		\centering
		\includegraphics[width=\linewidth]{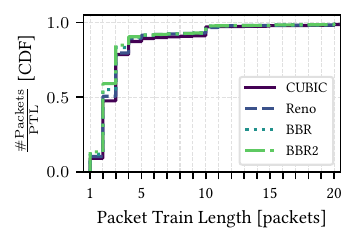}%
		\caption{\quiche}
	\end{subfigure}%
	\begin{subfigure}{0.33\linewidth}
		\centering
		\includegraphics[width=\linewidth]{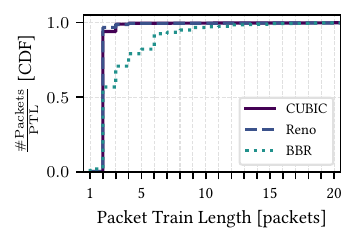}%
		\caption{\ngtcp}
	\end{subfigure}%
	\caption{Comparison of the \quic libraries \picoquic, \quiche and \ngtcp configured with different \acp{cca}. For each library, the top subfigure shows the inter-packet gaps while the bottom subfigure shows the packet train lengths (PTLs).}
    \Description[Comparison of the \quic libraries \picoquic, \quiche and \ngtcp configured with different \acp{cca}.]{Comparison of the \quic libraries \picoquic, \quiche and \ngtcp configured with different \acp{cca} (CUBIC, Reno, and BBR). The distributions of inter-packet gaps and packet train lengths are shown.} %
	\label{fig:baseline-impl-cca-comparison}
\end{figure*}%

For \tcptls and \ngtcp, more than \SI{99.9}{\percent} of the packets are in packet trains consisting of five packets or less, demonstrating consistent pacing behavior.
In contrast, \picoquic limits packet trains to five packets or fewer for only \SI{60}{\percent} of the packets, and \quiche achieves this for \SI{89}{\percent}.
After taking a closer look at the distributions, we see that \picoquic has \SI{3.6}{\percent} of packet trains, making up almost \SI{40}{\percent} of all packets, consisting of 16 or 17 packets.
We observed that those bursts are usually sent after a \SI{5}{\milli\second} idle period happening almost every \SI{10}{\milli\second}.
This behavior is visible with \cubic and \newreno as \ac{cca} but not present with \ac{bbr}.
\Cref{fig:baseline-impl-cca-comparison} compares the pacing behavior for different \acp{cca}.
The built-in optimization of \ac{bbr} in \picoquic to pace traffic can be fully utilized and traffic is close to perfectly spaced.
Only loss-based \acp{cca} result in bursty behavior in \picoquic.
This behavior is not visible with the other implementations.
For \quiche and \ngtcp, bursts are smaller when \cubic or \newreno are used.
However, both libraries do not reach the pacing behavior of \picoquic if \ac{bbr} is used, but the pacing behavior is similar to the baseline.
The \ac{bbr} implementation of \ngtcp even leads to an increase of loss by an order of magnitude.

The packet train length of \quiche is more evenly distributed across values between 6 and 20 and does not show any significant outliers.
Looking at \Cref{tab:baseline-metrics}, it is evident that the standard deviation of both metrics is highest for \quiche.
A patch for \quiche presented in \Cref{sec:evaluation-fq} unravels the origin of these observations, as they were not caused by the missing pacing.
\quiche itself does not pace packets but relies on a suitable \ac{qdisc} to schedule packets based on timestamps.

\begin{table}[tb]
    \centering
    \caption{Reached goodput and number of dropped packets during the baseline measurements. The bandwidth is limited to \SI{40}{\Mbps}. Packets are dropped due to a full buffer at the introduced bottleneck.}
    \label{tab:baseline-metrics}
    \sisetup{separate-uncertainty, table-number-alignment=right,retain-zero-uncertainty=true}
    \begin{tabular}{l S[table-format=3.2(3.2)] S[table-format=2.2(1.2)]}
        \toprule
        Implementation & {Dropped packets} & {Goodput [\si{\mega\bit\per\second}]} \\
        \midrule
        \quiche & \resultsBaselineQuicheDrops(\resultsBaselineQuicheDropsStd) & \resultsBaselineQuicheGoodput(\resultsBaselineQuicheGoodputStd) \\
        \picoquic & \resultsBaselinePicoquicDrops(\resultsBaselinePicoquicDropsStd) & \resultsBaselinePicoquicGoodput(\resultsBaselinePicoquicGoodputStd) \\
        \ngtcp & \resultsBaselineNgtcpDrops(\resultsBaselineNgtcpDropsStd) & \resultsBaselineNgtcpGoodput(\resultsBaselineNgtcpGoodputStd) \\
        \tcptls & \resultsBaselineTcptlsDrops(\resultsBaselineTcptlsDropsStd) & \resultsBaselineTcptlsGoodput(\resultsBaselineTcptlsGoodputStd) \\
        \bottomrule
    \end{tabular}
\end{table}

\subsection{Fair Queue}
\label{sec:evaluation-fq}

\newcommand{\resultsFqDefaultDrops}{687.15}
\newcommand{\resultsFqDefaultDropsStd}{338.12}
\newcommand{\resultsFqDefaultGoodput}{34.67}
\newcommand{\resultsFqDefaultGoodputStd}{0.64}
\newcommand{\resultsFqFqDrops}{1022.55}
\newcommand{\resultsFqFqDropsStd}{324.33}
\newcommand{\resultsFqFqGoodput}{33.64}
\newcommand{\resultsFqFqGoodputStd}{0.89}
\newcommand{\resultsFqSfDrops}{39.45}
\newcommand{\resultsFqSfDropsStd}{32.81}
\newcommand{\resultsFqSfGoodput}{31.15}
\newcommand{\resultsFqSfGoodputStd}{0.26}
\newcommand{\resultsFqFqsfDrops}{6.35}
\newcommand{\resultsFqFqsfDropsStd}{1.01}
\newcommand{\resultsFqFqsfGoodput}{31.06}
\newcommand{\resultsFqFqsfGoodputStd}{0.33}

\newcommand{\resultsFqDefaultGoodputFmt}{\resultsFqDefaultGoodput(\resultsFqDefaultGoodputStd)}
\newcommand{\resultsFqFqGoodputFmt}{\resultsFqFqGoodput(\resultsFqFqGoodputStd)}
\newcommand{\resultsFqDefaultDropsFmt}{\resultsFqDefaultDrops(\resultsFqDefaultDropsStd)}
\newcommand{\resultsFqFqDropsFmt}{\resultsFqFqDrops(\resultsFqFqDropsStd)}

\begin{figure}
	\centering
	\includegraphics[]{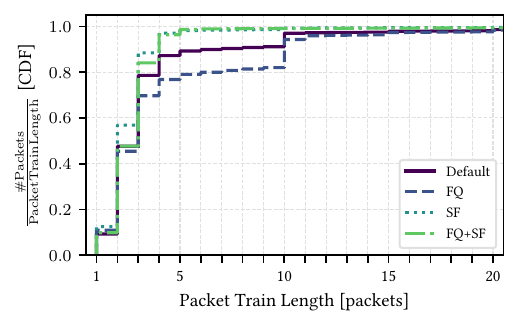}
	\caption{The impact of the \ac{fq} \ac{qdisc} on \quiche pacing. SF denotes our \quiche patch to change the behavior of the spurious loss detection. Without the patch, pacing can be drastically impacted in case of loss.}
	\label{fig:fq-trainsize-pp-cdf}
\end{figure}

To evaluate kernel-supported pacing, we use the \ac{fq} \ac{qdisc}, which integrates with \quiche's pacing mechanism to schedule packets based on timestamps.
With \ac{fq}, we surprisingly observe a performance decrease.
While the goodput worsens to \SIGoodputPlusStd{\resultsFqFqGoodputFmt}, the number of dropped packets increases to \SIPacketDropsPlusStd{\resultsFqFqDropsFmt}.
The overall stability of metrics decreased, resulting in an increased standard deviation for both metrics.
The tail of the packet train length distribution shown in \Cref{fig:fq-trainsize-pp-cdf} reveals that the number of packet trains with more than five packets increased.
A detailed analysis of the source code and individual loss events shows that after a packet is lost, \quiche periodically performs a rollback on the congestion window.
This happens multiple times in a row, leading to more packet loss.
We were able to observe periods of up to \SI{5}{\second} where the congestion window fluctuates between two values, causing a high number of packet loss.
The only scenario in which \cubic may roll back the congestion window is the detection of a spurious congestion event~\cite{rfc9438}.
While spurious loss is usually defined as receiving an \ac{ack} for a packet considered lost, \quiche also defines any loss that involves a number of packets lower than a specific threshold as spurious loss~\cite{quiche-loss}.
When \quiche detects packet loss after the start of the current congestion recovery period, it creates a congestion event and initiates a new recovery period, with the congestion controller state checkpointed beforehand.
If the acknowledged packet was sent after the recovery period began, \quiche's \cubic algorithm checks whether the increase in lost packets since the checkpoint is below a threshold and restores the checkpoint state if so, creating a risk of perpetual congestion window rollbacks.
An example of this behavior can be seen in Appendix \Cref{fig:quiche-spurious}.
As observed, pacing worsens this issue by reducing packet loss per cycle, which increases the chances of rollbacks, whereas \quiche recovers more quickly in the baseline measurements because of larger bursts of loss.
Although this mechanism was originally introduced to improve performance~\cite{cloudflare-issue}, we deactivate it for all further measurements, as it causes aggressive behavior and worse performance in our setup.

Comparing the pacing of the \quiche baseline with the \ac{fq} configuration and the previously mentioned mechanism disabled, we observe similar pacing during large parts of the connection.
However, with \ac{fq}, packet trains longer than five packets are rare while they make up over \SI{10}{\percent} of the packets in the baseline.
Due to the ACK-clocking effect, packets are paced during large parts of the connection with both configurations.
Only after a congestion event, we observe that ACK-clocking does not lead to a stable pacing behavior without the \ac{fq} \ac{qdisc}.
This happens because the congestion window is decreased, but \acp{ack} are still arriving at the old rate.
This offset leads to delayed bursts, which are not present when using \ac{fq}.

\subsection{GSO}
\label{sec:evaluation-gso}

\newcommand{\resultsGsoGsoenabledDrops}{6.35}
\newcommand{\resultsGsoGsoenabledDropsStd}{1.01}
\newcommand{\resultsGsoGsoenabledGoodput}{31.06}
\newcommand{\resultsGsoGsoenabledGoodputStd}{0.33}
\newcommand{\resultsGsoGsodisabledDrops}{160.80}
\newcommand{\resultsGsoGsodisabledDropsStd}{39.17}
\newcommand{\resultsGsoGsodisabledGoodput}{31.71}
\newcommand{\resultsGsoGsodisabledGoodputStd}{0.08}
\newcommand{\resultsGsoGsopacedDrops}{166.20}
\newcommand{\resultsGsoGsopacedDropsStd}{41.00}
\newcommand{\resultsGsoGsopacedGoodput}{31.71}
\newcommand{\resultsGsoGsopacedGoodputStd}{0.07}

\begin{figure}
	\centering
	\includegraphics[]{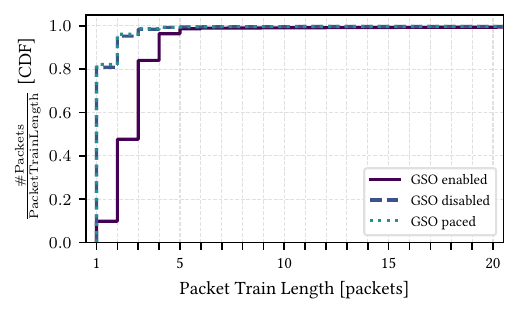}
    \caption{The impact of \ac{gso} on the pacing behavior of \quiche. Combining multiple \quic packets into one buffer reduces system calls but increases the burstiness. \ac{gso}-paced is based on a kernel patch that allows senders to provide a pacing rate with the \ac{gso} buffer.}
	\label{fig:gso-trainsize-pp-cdf}
\end{figure}

\ac{gso} can significantly reduce \ac{cpu} overhead for packet processing in \quic.
However, its batching nature leads to bursts as shown in \Cref{fig:gso-trainsize-pp-cdf}.
Pacing single packets with \acp{qdisc}, as introduced in \Cref{sec:evaluation-fq}, remains possible with batching methods like \texttt{sendmmsg()}, but not with \ac{gso}.
The \ac{gso} buffer size, controlled by the \quic implementation, directly influences this burstiness.
While \ac{gso} reduces \ac{cpu} load, the resulting bursty traffic might cause a performance degradation.

To mitigate the burstiness introduced by \ac{gso}, two approaches can be considered.
The easier approach is to send smaller \ac{gso} bursts and to pace the gaps between them.
This can be done by adjusting the \ac{gso} buffer size calculation in the \quic implementation.
However, this approach does not fully utilize the advantages of \ac{gso} and requires a trade-off between \ac{cpu} load and burstiness.
The second approach is to pace individual packets within the \ac{gso} buffer at the kernel level.
Our evaluation focuses on the second approach, as it allows large buffers to be passed to the kernel, maximizing the benefits of \ac{gso} while enabling the kernel to pace packets individually.
The trade-off here, however, is that this requires modifications to the kernel.

\begin{table}[bt]
    \centering
    \caption{Reached goodput and number of dropped packets during the \ac{gso} measurements. The bandwidth is limited to \SI{40}{\Mbps}. Packets are dropped due to a full buffer at the introduced bottleneck.}
    \label{tab:gso-metrics}
    \sisetup{separate-uncertainty, table-number-alignment=right}
    \begin{tabular}{l S[table-format=3.2(3.2)] S[table-format=2.2(1.2)]}
        \toprule
        GSO & {Dropped packets} & {Goodput [\si{\mega\bit\per\second}]} \\
        \midrule
        enabled & \resultsGsoGsoenabledDrops(\resultsGsoGsoenabledDropsStd) & \resultsGsoGsoenabledGoodput(\resultsGsoGsoenabledGoodputStd) \\
        disabled & \resultsGsoGsodisabledDrops(\resultsGsoGsodisabledDropsStd) & \resultsGsoGsodisabledGoodput(\resultsGsoGsodisabledGoodputStd) \\
        paced & \resultsGsoGsopacedDrops(\resultsGsoGsopacedDropsStd) & \resultsGsoGsopacedGoodput(\resultsGsoGsopacedGoodputStd) \\
        \bottomrule
    \end{tabular}
\end{table}

We implemented a kernel patch for \ac{gso} pacing based on a proposal by Willem de Bruijn~\cite{gso-pacing-patch}.
The patch was adapted for easier integration with \quic libraries, enabling the kernel to accept a pacing rate in bytes per second for each \ac{gso} buffer.
\Cref{fig:gso-trainsize-pp-cdf} shows the results.
We compare three variants of \quiche with paced traffic:
\ac{gso} is disabled in the first measurement, \ac{gso} is enabled in the second, and \ac{gso} is enabled with our kernel patch, allowing the kernel to pace packets within each \ac{gso} batch, in the third.
The results show that paced \ac{gso} achieves pacing behavior similar to \ac{gso}-disabled configurations.
Over \SI{80}{\percent} of packets are sent outside a packet train, indicating effective pacing.
However, \Cref{tab:gso-metrics} shows that packet loss increases to nearly ten times that of unpaced \ac{gso}.
This additional loss occurs only at the end of the slow start phase.
We attribute this behavior to \hystart, a modification to the slow start phase to exit early based on \ac{rtt} increases~\cite{rfc9406}.
With standard \ac{gso}, the bursty traffic fills the bottleneck buffer, causing a rapid \ac{rtt} increase that causes an early exit from slow start.
In contrast, paced \ac{gso} and \ac{gso}-disabled configurations exhibit smoother traffic patterns, leading to a slower \ac{rtt} increase that does not trigger early slow start exits.
Adjusting the parameters of \hystart based on the performed pacing could mitigate this issue.

\subsection{ETF and Hardware Offloading}
\label{sec:evaluation-etf}

We now evaluate the \ac{etf} \ac{qdisc}, which offers support for \launchtime to offload pacing to the \ac{nic}.
For that, we compare \ac{fq}, \ac{etf}, and \ac{etf} with \launchtime, all with paced \ac{gso}.
As described in \Cref{sec:background}, \ac{etf} requires a \textit{delta} parameter suitable for the used hardware.
\citet{9910175} identified \SI{175}{\micro\second} to be most suitable on a system similar to ours, but they argue that a higher value could reduce packet drops.
To be a bit more conservative, we chose a \textit{delta} of \SI{200}{\micro\second}.

The distributions of the inter-packet gaps and the packet train lengths for the three measurements don't show any significant differences.
As the implementations of both \acp{qdisc} do not differ much, this is expected.

To evaluate if the \launchtime feature improves pacing precision, we compare expected and actual send timestamps for all packets. 
As \ac{gso} introduces extra complexity and lowers the amount of samples for this metric, the measurements are done without \ac{gso}.
The expected timestamp together with the \quic packet number is logged by the \quiche server while the actual timestamp and the packet number are retrieved from the packet capture of the sniffer.
While the actual difference is not a suitable metric here, the standard deviation, from now on referenced as precision, of the differences is.
It is independent of the mean average, which is not meaningful as the clocks of server and sniffer are not synchronized in our setup.

The precision of \ac{etf} with and without hardware offloading is \SI{0.27}{\milli\second} and \SI{0.28}{\milli\second} respectively.
As the behavior and precision do not differ from the \ac{etf} \ac{qdisc} without hardware offloading, we do not consider the \launchtime feature to be beneficial for pacing.
Others have also experienced no increase in performance when using \ac{etf} with \launchtime~\cite{10150312}.
To our surprise, the precision of \ac{fq} is \SI{0.12}{\milli\second}, which is better than the \ac{etf} implementations.
The baseline measurements without any \ac{qdisc} show the worst precision with \SI{0.94}{\milli\second}.
This is also expected, as the kernel does not process the timestamps of the packets in this case.

\section{Related Work}
\label{sec:related-work}

Pacing for TCP was already investigated in 2000 by \citeauthor{Aggarwal2000pacing}~\cite{Aggarwal2000pacing}.
They showed that pacing can have advantages in many scenarios but might also negatively impact the performance.
More recent studies confirmed many of these findings~\cite{grazia2021ontheblock} and focused on use cases, \eg integrated into \ac{cca}~\cite{Cardwell2017bbr,song2021bbr2,zeynali2024bbr} or video streaming scenarios~\cite{spang2023samy}.
The ongoing importance of pacing in transport protocols is further highlighted by recent discussions within the IETF, where \citeauthor{welzl-iccrg-pacing-02}~\cite{welzl-iccrg-pacing-02} provide an overview of pacing mechanisms and their role in improving network performance.
Its existence underscores that effective pacing, particularly for user-space protocols like \quic, remains a relevant area of research.

Differences and the performance of \quic compared to the existing stack have been evaluated by different works~\cite{jaeger2023quic,KEMPF202490,yang2020quicnicoffloading,tyunyayev2022picoquichighspeed,yu2021quicproductionperformance,megyesi2016howquickisquic,shreedhar2022quicperformancewebandstorage,wolsing2019performance,bauer2023tcpandquic}.
However, they mostly focus on the reachable goodput, page load times, or offloading features without considering pacing behavior of implementations.

In 2020, \citeauthor{marx2020implementationdiversity}~\cite{marx2020implementationdiversity} compared different \quic implementations based on their source code.
They find that only 8 out of 15 libraries implement pacing.
Other research argues that \quic has advantages due to pacing~\cite{de2018optimizing,zhang2024quickenough,yu2017quicmeetstcp}.
However, the quality of the actual pacing is not evaluated.
Fastly~\cite{fastly-gso} and Cloudflare~\cite{cloudflare-gso} have discussed the effects of optimizations on performance, \eg \ac{gso}, but also mentioned drawbacks regarding pacing and suggest further investigation.
In contrast, \citeauthor{manzoor2019wifi}~\cite{manzoor2019wifi} explicitly prevent pacing to improve \quic performance in WiFi.
While the increased burstiness improves their results, they did not evaluate inter-packet gaps and the actual pacing behavior in more detail.

We show that even though pacing might be implemented, the interplay of \quic as user-space implementation with the kernel and offloading features can have a drastic impact.
These effects have to be considered in the future and configurations should be adapted based on the use case.

\section{Conclusion}
\label{sec:conclusion}

In this work, we evaluated the pacing behavior of three \quic libraries, Cloudflare \quiche, \ngtcp and \picoquic.
We shed light on the impact of different \acp{cca} and functionalities offered by the kernel.
Our results show that pacing behavior differs widely between libraries.
While \picoquic offers a \ac{bbr} implementation which evenly spaces packets, other libraries show advantages using CUBIC.
\quiche can make use of the \ac{fq} \ac{qdisc} for pacing but also \ac{gso} to reduce context switches between the library in user-space and the kernel.
We show that both functionalities reach the targeted goal, but while \ac{gso} reduces the sending overhead, it drastically impacts packet spacing.
Our adapted kernel patch for paced \ac{gso}, optimized for easy integration into \quic implementations, combines the efficiency of batching to reduce the overhead of system calls while keeping the pacing behavior.
Hardware offloading does not show relevant improvements and reduces pacing precision as intended by the library. %

Finally, our work shows that pacing in \quic is possible and can be achieved using different approaches.
Accurate pacing can be entirely done from user-space as shown by \picoquic with \ac{bbr} or with the help of kernel functionality such as the \ac{fq} \ac{qdisc} or when using \ac{gso}.
However, the behavior differs between implementations and approaches.
Depending on the application use case, \eg video streaming, real-time communications, or web access, different pacing strategies or even no pacing at all might be beneficial.
Our evaluation offers important insights for a better understanding of pacing in \quic, contributing to the ongoing development and successful usage in different scenarios.

\section*{Acknowledgment}
\label{sec:acknowledgment}

We thank the anonymous reviewers and our shepherd for their valuable feedback.
This work was supported by the EU’s Horizon 2020 programme as part of the projects SLICES-PP (10107977) and GreenDIGIT (101131207),
by the German Federal Ministry of Education and Research (BMBF) under the projects 6G-life (16KISK002) and 6G-ANNA (16KISK107),
and by the German Research Foundation (HyperNIC, CA595/13-1).

\bibliographystyle{ACM-Reference-Format}
\bibliography{paper.bib}

\appendix

\section{Spurious Loss Behavior: quiche}
\begin{figure}[h]
	\includegraphics[]{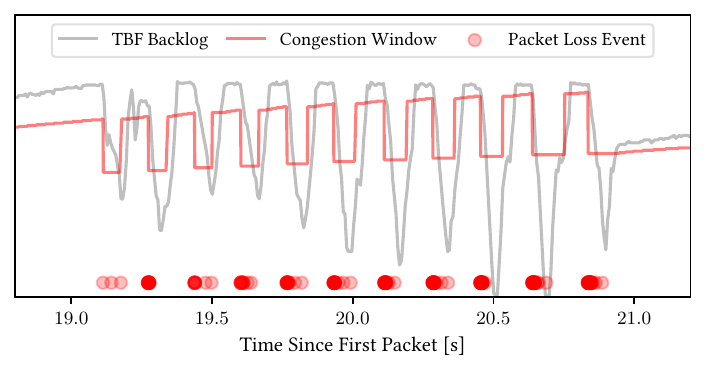}
	\caption{Example of the \quiche behavior in case of spurious loss. The current implementation can lead to perpetual congestion window rollbacks.}
	\label{fig:quiche-spurious}
\end{figure}

\Cref{fig:quiche-spurious} shows the behavior of the original \quiche implementation in case of spurious loss.  
The behavior is explained in more detail in \Cref{sec:evaluation-fq}.
It is also discussed in a GitHub issue~\cite{cloudflare-issue}.
It can drastically impact pacing behavior and performance.
Therefore, we deactivate the behavior by patching \quiche.

\section{Artifacts}
The source code of our measurement framework introduced in \Cref{sec:approach} is published on GitHub~\cite{quic-pacing-paper-repo}.
As explained, it is an extension to the framework published by \citeauthor{jaeger2023quic}~\cite{jaeger2023quic}.
Furthermore, this repository contains the configurations for all measurements and all used patches for \quiche and paced \ac{gso}.
The framework and configurations can be used to reproduce results and study further configurations or libraries.

Besides these artifacts to allow measurements, the collected data from our measurements presented in this work is also published in the same repository.
The data contains detailed logs, packet captures and additional information from the involved systems.
By using the included evaluation scripts, the data can be analyzed and visualized.

\label{lastpage}

\end{document}